\documentclass[aps,prd,nofootinbib,amsmath,amssymb,superscriptaddress,twocolumn,10pt]{revtex4}
\usepackage{amsmath}
\usepackage{amsfonts}
\usepackage{graphicx}
\usepackage{dcolumn}
\usepackage{bbm}
\usepackage{amssymb}
\usepackage{latexsym}
\usepackage{CJK}
\usepackage{titlesec}
\usepackage{color}
\usepackage[colorlinks=true, linkcolor=red, citecolor=blue]{hyperref}
\usepackage{appendix}
\let\sb=_ \catcode`\_=\active \def_#1{\ensuremath \sb{\rm#1}}

\begin{document}

 \bibliographystyle{unsrt}

\title{Constraints on brane inflation after Planck 2015: Impacts of the latest local measurement of the Hubble constant}

\author{Rui-Yun Guo}
\affiliation{Department of Physics, College of Sciences, Northeastern University,
Shenyang 110819, China}
\author{Lei Zhang}
\affiliation{Department of Physics, College of Sciences, Northeastern University,
Shenyang 110819, China}
\author{Jing-Fei Zhang}
\affiliation{Department of Physics, College of Sciences, Northeastern University,
Shenyang 110819, China}
\author{Xin Zhang\footnote{Corresponding author}}
\email{zhangxin@mail.neu.edu.cn}
\affiliation{Department of Physics, College of Sciences,
Northeastern University, Shenyang 110819, China}
\affiliation{Center for High Energy Physics, Peking University, Beijing 100080, China}

\begin{abstract}
We investigate the observational constraints on three typical brane inflation models by considering the latest local measurement of the Hubble constant in the global fit. We also employ other observational data, including the Planck 2015 CMB data, the BICEP2/Keck Array B-mode data, and the baryon acoustic oscillations data, in our analysis. Previous studies have shown that the addition of the latest local $H_{0}$ measurement favors a larger spectral index, and can exert a significant influence on the model selection of inflation. In this work, we investigate its impacts on the status of brane inflation models. We find that, when the direct $H_{0}$ measurement is considered, the prototype model of brane inflation is still in good agreement with the current observational data within the $2\sigma$ level range. For the KKLMMT model, the consideration of the $H_{0}$ measurement allows the range of the parameter $\beta$ to be amplified to ${\cal O}(10^{-2})$, which slightly alleviates the fine-tuning problem. For the IR DBI model, the addition of the $H_{0}$ measurement does not provide a better fit. These results show that the consideration of the new $H_{0}$ prior can exert a considerable influence on the brane inflation models. At last, we show that, when $\beta \lesssim 1.1$, the equilateral non-Gaussianity in the IR DBI inflation model is compatible with the current CMB data at the 1$\sigma$ level.
\end{abstract}

\maketitle

\renewcommand{\thesection}{\arabic{section}}
\renewcommand{\thesubsection}{\arabic{subsection}}
\titleformat*{\section}{\flushleft\bf}
\titleformat*{\subsection}{\flushleft\bf}

\section{Introduction}

Inflation refers to a period of accelerated expansion of the very early universe in cosmology. It can successfully resolve the horizon problem, the flatness problem, and the monopole problem of the standard hot big-bang cosmology~\cite{Guth: 1981,Albrecht: 1982,Linde: 1982}. Inflation provides a leading paradigm to explain the origin of the primordial density perturbations and the primordial gravitational waves. Various phenomenological models of inflationary universe have been proposed, but the microscopic mechanism of inflation is still needed to be explained from some fundamental theory of physics. By far, string theory is still viewed as the most promising candidate for a fundamental theory, and thus it is natural to expect that a theoretical scenario of inflation could be realized within the framework of the superstring theory. Brane inflation \cite{Dvali:1998pa,HenryTye:2006uv} is just such a typical inflation scenario based on string theory.

Because the primordial density perturbations generated by inflation can leave rich imprints in the anisotropies of the cosmic microwave background (CMB) and the large-scale structure formation, these observations can be naturally employed as useful tools to examine different inflation models. A popular method for examining the status of an inflation model is to compare the predictions of the inflation model with current observations (see, e.g., Refs.~\cite{Liu:2010fm,Li:2013qga,Tsujikawa:2013ila,Wan:2014fra,Cai:2015soa,Huang:2015xda,Cai:2015gla,Huang:2015gca,Huang:2015cke,Li:2016awk,Xia:2014tda,Liu:2014tda,Tram:2016rcw,Xu:2016kwz,Cai:2016ngx,Gerbino:2016sgw,Linde:2014nna,Zhang:2017epd,Guo:2017qjt,Obied:2017tpd,Ni:2017jxw} and references therein).

In Refs.~\cite{Zhang:2017epd,Guo:2017qjt}, we have revisited the constraints on some selected slow-roll inflation models. Our results show that, when the latest local measurement of the Hubble constant is considered, inflation models with a concave potential are more favored than those with a convex potential; for the natural inflation model and the $V\propto \phi^{2}$ model with a convex potential, they are totally excluded at more than $2\sigma$ level; the Starobinsky $R^{2}$ inflation model and the $V\propto \phi$ model are marginally favored at around $2\sigma$ level; the $V\propto \phi^{2/3}$ model with a concave potential is located in the $1\sigma$ region; and the spontaneously broken SUSY inflation model is now the most favored model. Compared with the results of Huang et al.~\cite{Huang:2015cke}, in Refs.~\cite{Zhang:2017epd,Guo:2017qjt} we conclude that the addition of the latest $H_{0}$ measurement ($H_{0} = 73.00\pm 1.75$ km s $^{-1}$ Mpc$^{-1}$) can further exclude (or favor) some special slow-roll inflation models. References~\cite{Tram:2016rcw,Santos:2017alg} also conclude that the direct $H_{0}$ measurement indeed provides a large value of the spectral index, which further affects the inflation model selection. In this paper, we will reexamine the status of brane inflation models in light of the current observational data, after the Planck 2015 results, including the latest $H_{0}$ measurement.

Brane inflation can be successfully realized in string theory, and this theory can provide a UV completion to any effective field theory. In the brane inflationary scenario, the inflation is driven by the motion of a dynamic brane towards an antibrane. The distance between the brane and the antibrane in the extra compactified dimensions is interpreted as the inflaton. The brane inflationary scenario can be realized via two effective mechanisms, which are the slow-roll inflation and the Dirac-Born-Infeld (DBI) inflation~\cite{Bean:2007eh}. For the slow-roll mechanism, we consider the prototype model~\cite{Quevedo:2002xw} and the KKLMMT model~\cite{Kachru:2003sx}. For the DBI mechanism, we consider the popular infrared (IR) DBI model. For other extensive studies on the brane inflation models, see, e.g, Refs.~\cite{Santos:2017alg,Bean:2007eh,Firouzjahi:2005dh,Huang:2006ra,Huang:2006zu,Zhang:2006cc,Baumann:2006cd,Baumann:2007ah,Baumann:2008kq,Ma:2008rf,Ma:2013xma,Gangopadhyay:2016qqa}.

In this paper, we consider the same observational data as used in Ref.~\cite{Guo:2017qjt}. They are the Planck 2015 CMB data including the temperature and polarization power spectra at whole multipoles as well as the lensing reconstruction (Planck)~\cite{Ade:2015xua}, the BICEP2/Keck Array B-mode data (BK)~\cite{Ade2016}, the four baryon acoustic oscillations data (BAO)~\cite{Beutler:2011hx,Ross:2014qpa,Gil-Marin:2015nqa}, and the latest $H_{0}$ measurement~\cite{Riess:2016jrr}. The latest measurement value of $H_{0}$ has a reduction of the uncertainty from $3.3\%$ to $2.4\%$ with the Wide Field Camera 3 on the Hubble Space Telescope. Thus it is expected that it can lead to significant impacts on the constraints of the brane inflation models. For more detailed introduction for these data, we refer the reader to Ref.~\cite{Guo:2017qjt}.

An important aspect that is worth mentioning concerns the tension between the direct $H_{0}$ measurement and the Planck observation. To avoid the extra effect of the tension on cosmological models, we consider the additional parameter $N_{eff}$ in the cosmological model because the addition of the parameter $N_{eff}$ is considered to be a good scheme to relieve the tension between the direct $H_{0}$ measurement and the Planck observation~\cite{Zhang:2017epd,Guo:2017qjt,Riess:2016jrr,Zhang:2014dxk,Zhang:2014ifa,DiValentino:2016ucb,Benetti:2017gvm}. In fact, in the literature it is shown that the tension could also be relieved to some extent by considering a dynamical dark energy~\cite{Li:2013dha,Qing-Guo:2016ykt,DiValentino:2017zyq,DiValentino:2017iww,Barenboim:2017sjk} or light sterile neutrinos~\cite{Zhang:2014dxk,Li:2014kla,Zhang:2014ifa,Zhang:2014nta,Vagnozzi:2017ovm,Feng:2017nss,Zhao:2017urm,Hamaguchi:2017ihw,Feng:2017mfs,Zhao:2017jma}.

In this paper, we will examine these typical brane inflation models with the current observations. We consider two combinations of observational data: the Planck+BK+BAO data combination and the Planck+BK+BAO+$H_{0}$ data combination. The aim of this work is to investigate how the status of these brane inflation models is affected by the latest local measurement of the Hubble constant. The organization of this paper is as follows. In Sec.~\ref{sec.2}, we introduce the brane inflation models studied in this paper, which are the prototype model, the KKLMMT model, and the IR DBI model. In Sec.~\ref{sec.3}, we compare the theoretical predictions of the brane inflation models with the observational data. The observational constraints on the $\Lambda$CDM+$r$+$d n_{\rm s}/d \ln k$+$N_{\rm eff}$ model are obtained by using the Planck+BK+BAO data and the Planck+BK+BAO+$H_{0}$ data. In the last section, we give conclusions for the impacts of the latest direct $H_{0}$ measurement on the status of these brane inflation models.

\section {Models and methodology }\label{sec.2}

The brane inflation models can offer a large number of observational signatures \cite{Dvali:1998pa,HenryTye:2006uv}. In the following, some important information shall be given for the prototype model, the KKLMMT model, and the IR DBI model.

We first focus on a toy model of brane inflation, which is the prototype of brane inflation. In this scenario, a pair of $Dp$ and $\bar{D}p$-branes ($p\geq 3$) are put into the four large dimensions, and they are separated from each other in the extra six dimensions compactified. In fact, the toy model is not a realistic model, because it neglects an important problem, i.e., the distance between $Dp$ and $\bar{D}p$-branes would be larger than the size of the extra dimensional space when the inflaton slowly rolls, or $\eta$ is sufficiently small~\cite{Ma:2008rf,Huang:2006ra}. In general, the model is only considered as a warm-up exercise to confront cosmological observations.

The inflaton potential of this model is given by~\cite{Quevedo:2002xw,Ma:2008rf,Huang:2006ra,Ma:2013xma}
\begin{equation}\label{2.1}
  V=V_{0}\left(1-\frac{\mu^{n}}{\phi^{n}}\right),
\end{equation}
where $V_{0}$ denotes an effective cosmological constant on the brane. The second term in Eq.~(\ref{2.1}) provides the attractive force between the brane and the anti-brane. Because the transverse dimension $d_{\bot}=9-p\leq 6$, the parameter $n=d_{\bot}-2 \leq 4$. The inflaton $\phi$ is related to the $e$-folding number $N_{e}$ before the end of inflation~\cite{Ma:2008rf,Huang:2006ra,Ma:2013xma},
\begin{equation}\label{2.2}
  \phi=[N_{e}M_{pl}^{2}\mu^{n}n(n+2)]^{1/(n+2)},
\end{equation}
where $M_{pl}$ is the reduced Planck mass. By calculation, the slow-roll parameters of this model can be written as~\cite{Ma:2008rf,Huang:2006ra,Ma:2013xma}
\begin{equation}\label{2.3}
  \epsilon =\frac{n^{2}}{2(n(n+2))^{\frac{2(n+1)}{n+2}}}\left(\frac{\mu}{M_{pl}}\right)^{\frac{2n}{n+2}}N_{e}^{-\frac{2(n+1)}{n+2}},
\end{equation}
\begin{equation}\label{2.4}
  \eta=-\frac{n+1}{n+2}\frac{1}{N_{e}},
\end{equation}
\begin{equation}\label{2.5}
  \xi^{2}=\frac{n+1}{n+2}\frac{1}{N_{e}^{2}}.
\end{equation}

Next, we focus on a realistic brane inflation model---the KKLMMT model, which is derived from the type IIB string theory. This model considers the highly warped spacetime compactified and the moduli stabilization~\cite{Ma:2008rf,Kachru:2003sx}. With the addition of a small number of $\bar{D}3$-branes, the vacuum is successfully lifted to a de Sitter state. Furthermore, an extra pair of $D3$-brane and $\bar{D}3$-brane can be added in a warped throat. When the $D3$-brane moves towards the $\bar{D}3$-brane at the bottom of the throat, inflation takes place. In this scenario, the warped spacetime can provide a sufficiently flat potential, which can successfully resolve the ``$\eta$ problem" in the brane inflation. For more detailed introduction of the KKLMMT model, see Refs.~\cite{Bean:2007eh,Baumann:2006cd,Huang:2006ra,Ma:2008rf,Ma:2013xma,Kachru:2003sx}.

The inflaton potential of the KKLMMT model is given by~\cite{Ma:2008rf,Ma:2013xma}
\begin{equation}\label{2.6}
  V=\frac{1}{2}\beta H^{2}\phi^{2}+\frac{64\pi^{2}\mu^{4}}{27}(1-\frac{\mu^{4}}{\phi^{4}}),
\end{equation}
where $H$ is the Hubble parameter, $\beta$ is the coupling between inflaton $\phi$ and space expansion. In the KKLMMT model, we take approximately $\beta \simeq$ constant, and $|\beta| \ll 1$. The inflaton $\phi$ can be described as~\cite{Firouzjahi:2005dh,Ma:2008rf,Huang:2006ra,Ma:2013xma}
\begin{equation}\label{2.7}
  \phi^{6}=24M_{pl}^{2}\mu^{4}m(\beta),
\end{equation}
in which
\begin{equation}\label{2.8}
  m(\beta)=\frac{e^{2\beta N_{e}}(1+2\beta)-(1+\frac{1}{3}\beta)}{2\beta(1+\frac{1}{3}\beta)}.
\end{equation}

Combining Eqs.~(\ref{2.6}) and (\ref{2.7}), the slow-roll parameters of the KKLMMT model can be written as~\cite{Ma:2008rf,Huang:2006ra,Ma:2013xma}
\begin{equation}\label{2.9}
  \epsilon =\frac{1}{48}\left(\frac{3}{2}\right)^{\frac{1}{2}}(\Delta^{2}_{\mathcal{R}})^\frac{1}{2}m(\beta)^{-\frac{5}{2}}(1+2\beta m(\beta))^{3},
\end{equation}
\begin{equation}\label{2.10}
  \eta=\frac{\beta}{3}-\frac{5}{6}\frac{1}{m(\beta)},
\end{equation}
\begin{equation}\label{2.11}
  \xi^{2}=\frac{5}{3}\frac{1}{m(\beta)}\left[\beta+\frac{1}{2m(\beta)}\right],
\end{equation}
where the amplitude of the primordial scalar power spectrum is $\Delta^{2}_{\mathcal{R}} \simeq 2.2\times 10^{-9}$~\cite{Ade:2015xua}.

For the above slow-roll inflationary scenario, we have the relations between the observational quantities ($n_{\rm s}$, $r$, and $dn_{\rm s}/d\ln k$) and the slow-roll parameters ($\epsilon$, $\eta$, and $\xi^{2}$) ~\cite{Planck:2013jfk},
\begin{equation}\label{2.12}
  n_{\rm s} = 1- 6\epsilon + 2\eta,
\end{equation}
\begin{equation}\label{2.13}
  r = 16 \epsilon,
\end{equation}

\begin{equation}\label{2.14}
 dn_{\rm s}/d\ln k = 16\epsilon \eta -24\epsilon^{2}-2\xi^{2},
\end{equation}
where $r$, $n_{\rm s}$, and $dn_{\rm s}/d\ln k$ denote the tensor-to-scalar ratio (the ratio of the tensor spectrum to the scalar spectrum), the scalar spectral index, and the running of the scalar spectral index, respectively.

At last, we focus on another type of brane inflation scenario---the IR DBI model~\cite{Chen:2004gc,Chen:2005ad,Giddings:2001yu}. In this model, the inflaton rolls from the IR side to the UV side of the B-throat. The rolling velocity of the brane is dependent on the speed limit of the warped spacetime~\cite{Bean:2007eh}. Such a warped spacetime can always be naturally generated in the extra dimensions with the stabilized string compactification. In this scenario, the inflaton with a non-standard kinetic term is not slowly rolling at all. Moreover, because the sound speed of inflaton $c_{s}<1$ ~\cite{Ma:2008rf,Cheng:2012je}, a large tilt of the tensor power spectrum ($n_{t}=-r/(8c_{s})$) can be derived in the IR DBI model.

The inflaton potential of the IR DBI model is given by
\begin{equation}\label{2.15}
  V=V_{0}-\frac{1}{2}\beta H^{2}\phi^{2},
\end{equation}
with a wide range of $\beta\in (0.1,10^{9})$ allowed in principle~\cite{Bean:2007eh}. By calculation, the spectral index and its running can be reduced by~\cite{Bean:2007eh}
\begin{equation}\label{2.16}
  n_{\rm s}-1=\frac{4}{N_{e}^{\rm DBI}}\frac{x^{2}+3x-2}{(x+1)(x+2)},
\end{equation}
\begin{equation}\label{2.17}
  dn_{\rm s}/d\ln k=\frac{4}{(N_{e}^{\rm DBI})^{2}}\frac{x^{4}+6x^{3}-55x^{2}-96x-4}{(x+1)^{2}(x+2)^{2}},
\end{equation}
where $x=(N_{e}^{\rm DBI}/N_{\rm c})^{8}$. $N_{e}^{\rm DBI}$ denotes the number of $e$-folds of relativistic roll inflation, and can be estimated as $N_{e}^{\rm DBI}\approx \frac{H\sqrt{\lambda_{\rm B}}}{\phi}-\beta^{-1}$, where $\lambda_{\rm B}=n_{\rm B}N_{\rm B}/(2\pi^{2})$ ($n_{\rm B}$ is the number of branes in B throat and $N_{\rm B}$ is the effective charge of B throat). $N_{\rm c}$ denotes the critical DBI $e$-folds at $k_{c}$, in which string phase transition happens, and $N_{\rm c}$ is expressed as $N_{\rm c}=\sqrt{6}\pi^{1/4}\frac{N_{\rm B}^{1/8}}{\beta^{1/2}}$. The total number of $e$-folds $N_{e}^{\rm tot}=N_{e}^{\rm DBI} +N_{e}^{\rm NR}$, where $N_{e}^{\rm NR}$ denotes the $e$-folding number of non-relativistic roll inflation. In this paper, we employ the best-fit values given by Ref.~\cite{Bean:2007eh}, i.e., $N_{\rm c}=35.7$, $k_{\rm c}=10^{-4.15}$ Mpc$^{-1}$, and $N_{e}^{\rm NR}=18.4$, and we consider $N_{\rm e}^{\rm tot}\in [50,60]$, i.e., $N_{\rm e}^{\rm DBI}\in [31.6,41.6]$. Since $N_{\rm e}^{\rm DBI}$ involves various microscopic parameters, its dependence on $\beta$ cannot be measured, and thus in this paper we do not consider its relationship to $\beta$. Therefore, using Eqs.~(\ref{2.16}) and (\ref{2.17}) could not constrain the parameter $\beta$. Actually, the parameter $\beta$ of the IR DBI model can be well constrained by the primordial non-Gaussianities.



Primordial non-Gaussianities are characterized by the parameter $f_{\rm NL}$. As shown conclusively in Refs.~\cite{Maldacena:2002vr,Acquaviva:2002ud}, the non-Gaussianities are predicted to be undetectably small for slow-roll models of inflation. Thus in the discussions of the next section, we do not show the constraints on non-Gaussianities of the prototype model and the KKLMMT model. But for the DBI models, significant non-Gaussianities can be generated due to a small sound speed $c_{s}$ during the DBI inflation.\footnote{In Refs.~\cite{Chen:2006nt,Seery:2005wm}, the general expression of $f_{\rm NL}$ was given, $f_{\rm NL}=\frac{35}{108}\left(\frac{1}{c_{\rm s}^{2}}-1\right)$, which indicates that large non-Gaussianities will arise in models with $c_{\rm s}\ll 1$. While for the conventional slow-roll models with $c_{\rm s}=1$ preferred, the non-Gaussianities cannot be observed.}

Primordial non-Gaussianities are related to the shape of primordial fluctuations. In general, the shape has three types, i.e., the local, equilateral and orthogonal ones. The IR DBI model predicts an almost negligible local non-Gaussianity $f^{\rm loc}_{NL}$, which is consistent with current observations. The other two non-Gaussianities of the equilateral and orthogonal shapes are given by~\cite{Chen:2006nt,Senatore:2009gt},
\begin{equation}\label{2.18}
 f_{NL}^{\rm eq}=-0.35\frac{1-c_{s}^{2}}{c_{s}^{2}},
\end{equation}
\begin{equation}\label{2.19}
  f_{NL}^{\rm orth}=0.032\frac{1-c_{s}^{2}}{c_{s}^{2}},
\end{equation}
where the superscript ``eq" corresponds to the equilateral shape, ``orth" corresponds to the orthogonal shape, and $c_{s}\simeq 3/(\beta N_{e}^{\rm DBI})$. For more studies of the constraints on non-Gaussianities of primordial fluctuations, see Refs.~\cite{Bean:2007eh,Chen:2005fe,Planck:2013jfk,Ade:2015lrj,Renaux-Petel:2015bja}.

In this paper, we consider the usual cosmological parameters in the standard cosmology, namely, the baryon density, $\Omega_{\rm b}h^{2}$, the cold dark matter density, $\Omega_{\rm c}h^{2}$, the ratio between the sound horizon size and the angular diameter distance at the last-scattering epoch, $\theta_{\rm MC}$, the optical depth to reionization, $\tau$, the spectral index of the primordial power spectra of scalar perturbations, $n_{\rm s}$, and the tensor-to-scalar ratio, $r$. The additional parameters include the running spectral index, $d n_{\rm s}/d \ln k$, and the effective number of the extra relativistic degrees of freedom, $N_{\rm eff}$. The priors of all these parameters follow the Planck collaboration~\cite{Ade:2015xua}. We set the sum of neutrino masses as $\sum m_{\nu}=0.06$ eV and the pivot scale as $k_{0} = 0.05$ Mpc$^{-1}$. We take the reasonable range of the number of $e$-folds, $N_{e} \in [50,60]$. We modify the CosmoMC code~\cite{Lewis:2002ah} to solve the Boltzmann equations and explore the cosmological parameter space.

\section {Results}\label{sec.3}

In this section, we first constrain the $\Lambda$CDM+$r$+$d n_{\rm s}/d \ln k$+$N_{\rm eff}$ model by using the Planck+BK+BAO data and the Planck+BK+BAO+$H_{0}$ data. With the Planck+BK+BAO+$H_{0}$ data, the results of the constraints on the parameters $r$, $n_{\rm s}$ and $d n_{\rm s}/d \ln k$ have been given in Table I of Ref.~\cite{Guo:2017qjt}, i.e.,
$$
\small
\begin{array}{c}
r < 0.074~(2\sigma),\\
\\
n_{\rm s}=0 .9781\pm0.0080~(1\sigma), \\
\\
d n_{\rm s}/d \ln k=0 .0010^{+0.0074}_{-0.0073}~(1\sigma).\\
\end{array}
$$
Now, we give the constraints on these parameters by using the Planck+BK+BAO data,
$$
\small
\begin{array}{c}
r < 0.075~(2\sigma),\\
\\
n_{\rm s}=0.9670\pm0.0087~(1\sigma), \\
\\
d n_{\rm s}/d \ln k=-0.0020\pm0.0077~(1\sigma).\\
\end{array}
$$
We find that the consideration of the latest direct $H_{0}$ measurement prefers a larger spectral index $n_{\rm s}$ with a red tilt. No running of the spectral index is favored in both cases.

In the following, we will discuss the impacts of the latest measurement of $H_{0}$ on these typical brane inflation models. We examine whether predictions of the brane inflation models are consistent with the constraint results derived from current observations.

\subsection* {3.1 The prototype model}

\begin{figure*}[ht!]
\begin{center}
\includegraphics[width=7.6cm]{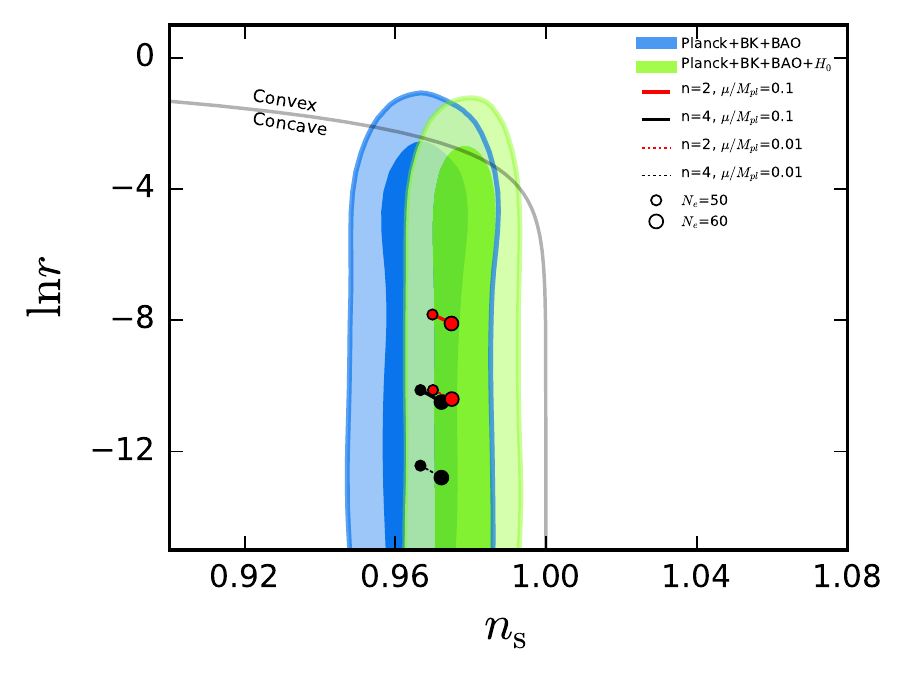}
\includegraphics[width=8.0cm]{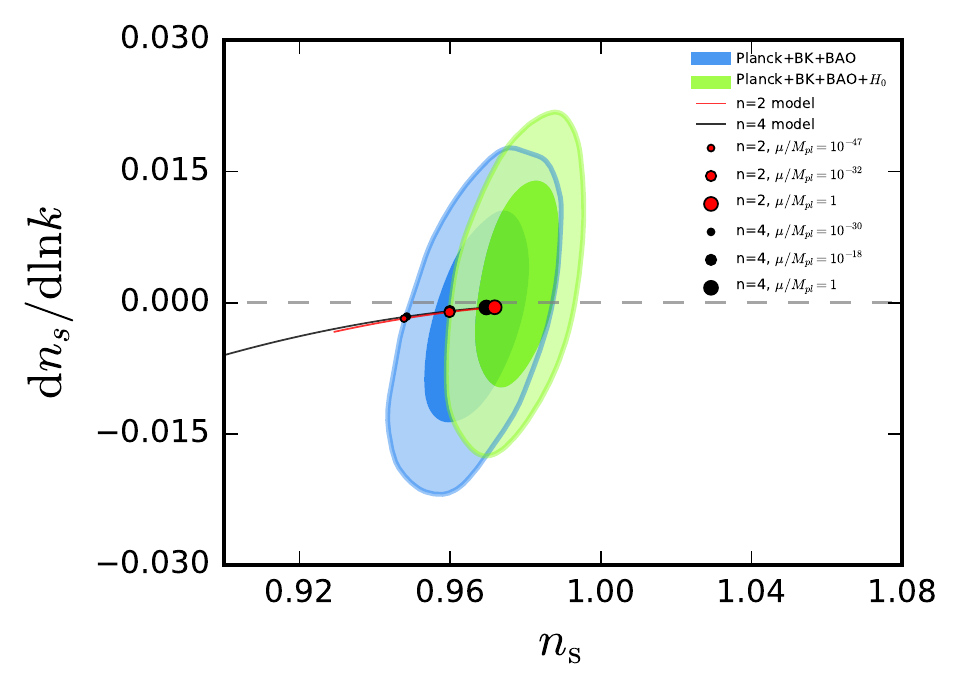}
\end{center}
\caption{Constraints ($1\sigma$ and $2\sigma$) on the $\Lambda$CDM+$r$+$d n_{\rm s}/d \ln k$+$N_{\rm eff}$ model in the $n_{\rm s}$--$r$ (\emph{left}) and $n_{\rm s}$--$d n_{\rm s}/d \ln k$ (\emph{right}) plane by using the Planck+BK+BAO and Planck+BK+BAO+$H_{0}$ data, compared to the theoretical predictions of the toy model. }
\label{fig1}
\end{figure*}

We plot the two-dimensional contours ($1\sigma$ and $2\sigma$) for $n_{\rm s}$ and $r$ by using the Planck+BK+BAO and Planck+BK+BAO+$H_{0}$ data, compared to the theoretical predictions of the toy model in the left panel of Fig.~\ref{fig1}. For a convenient display, we use $\ln r$ instead of $r$. Since the parameter $n\leq 4$ in theory, we focus on two cases of $n=2$ (red lines) and $n=4$ (black lines). We take $\mu/M_{pl}=0.1$ (solid lines) and $\mu/M_{pl}=0.01$ (dashed lines), respectively. The values between small and big dots correspond to the number of $e$-folds of $N_{e}\in [50,60]$. The gray line ($\eta =0$) can distinguish between the convex potential ($\eta >0$) and the concave potential ($\eta <0$). We see that most of the two-dimensional contours lies within the concave potential region, which indicates strong evidence for supporting the concave potential.

Comparing the blue and green contours in the left panel of Fig.~\ref{fig1}, we see that the addition of the direct $H_{0}$ measurement indeed favors a larger value of $n_{\rm s}$ [changing from $0.9670\pm0.0087$ ($1\sigma$) to $0.9781 \pm 0.0080$ ($1\sigma$)]. With the Planck+BK+BAO data, the predictions of both the $n=2$ model and the $n=4$ model lie within the $1\sigma$ region, indicating that the toy model is consistent with the Planck+BK+BAO data in both cases. When the direct $H_{0}$ measurement is included in the data combination, a part range of the predictions of $n=4$ model locates outside the $1\sigma$ region (but still inside the $2\sigma$ region). This result shows that the toy model is slightly worse favored with the addition of the direct $H_{0}$ measurement. But, overall, the predictions of the toy model are consistent with the observational constraints on $n_{\rm s}$ and $r$.

In the right panel of Fig.~\ref{fig1}, we give two-dimensional contours ($1\sigma$ and $2\sigma$) for $n_{\rm s}$ and $d n_{\rm s}/d \ln k$ by using the Planck+BK+BAO and Planck+BK+BAO+$H_{0}$ data, compared to the theoretical predictions of the toy model. The red and black lines are the predictions of the $n=2$ and $n=4$ models for different values of $\mu/M_{pl}$, respectively. We see that these two cases are almost overlapped, and parts of them locate inside the observationally allowed regions. By calculation, we figure out the allowed values of $\mu/M_{pl}$ that make the lines locate inside the $2\sigma$ regions. With the Planck+BK+BAO data, the ranges of $\mu/ M_{\rm pl}$ are required to be $[10^{-47},1]$ for the $n=2$ case and $[10^{-30},1]$ for the $n=4$ case. With the Planck+BK+BAO+$H_{0}$ data, the ranges are reduced to $[10^{-32},1]$ and $[10^{-18},1]$.

According to the approximate relationship between the amplitude of inflation and the parameter $\mu$~\cite{Ma:2013xma},
 \begin{equation}\label{3.1}
   \frac{V^{\frac{1}{4}}}{M_{\rm pl}}=\left(24\pi^{2} \Delta^{2}_{\mathcal{R}} \frac{n^{2}}{2(n(n+2))^{\frac{2(n+1)}{n+2}}}\left(\frac{\mu}{M_{\rm pl}}\right)^{\frac{2n}{n+2}}N_{e}^{-\frac{2(n+1)}{n+2}}\right)^{\frac{1}{4}},
 \end{equation}
and our estimation of $\mu/M_{pl}$, we obtain the ranges of the energy scale of inflation: $[1.8\times 10^{4},7.8\times 10^{15}]$ GeV ($n=2$) and $[6.8\times 10^{5},5.4\times 10^{15}]$ GeV ($n=4$) with the Planck+BK+BAO data, and $[9.0\times 10^{7},7.8\times 10^{15}]$ GeV ($n=2$) and $[6.1\times 10^{9},5.4\times 10^{15}]$ GeV ($n=4$) with the Planck+BK+BAO+$H_{0}$ data. These derived ranges belong to $[10^{3},10^{16}]$ GeV, indicating that these results are reasonable for current inflation theory. Given the above discussion, we reconfirm that the predictions of the toy model are still consistent with the current observational data.

\subsection* {3.2 The KKLMMT model}

\begin{figure*}[ht!]
\begin{center}
\includegraphics[width=7.6cm]{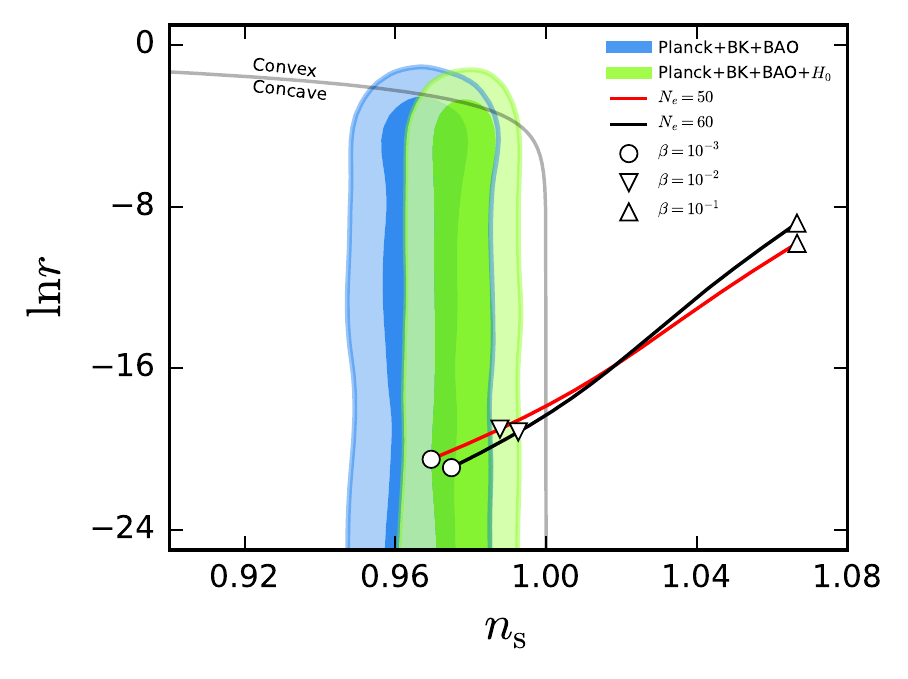}
\includegraphics[width=8.0cm]{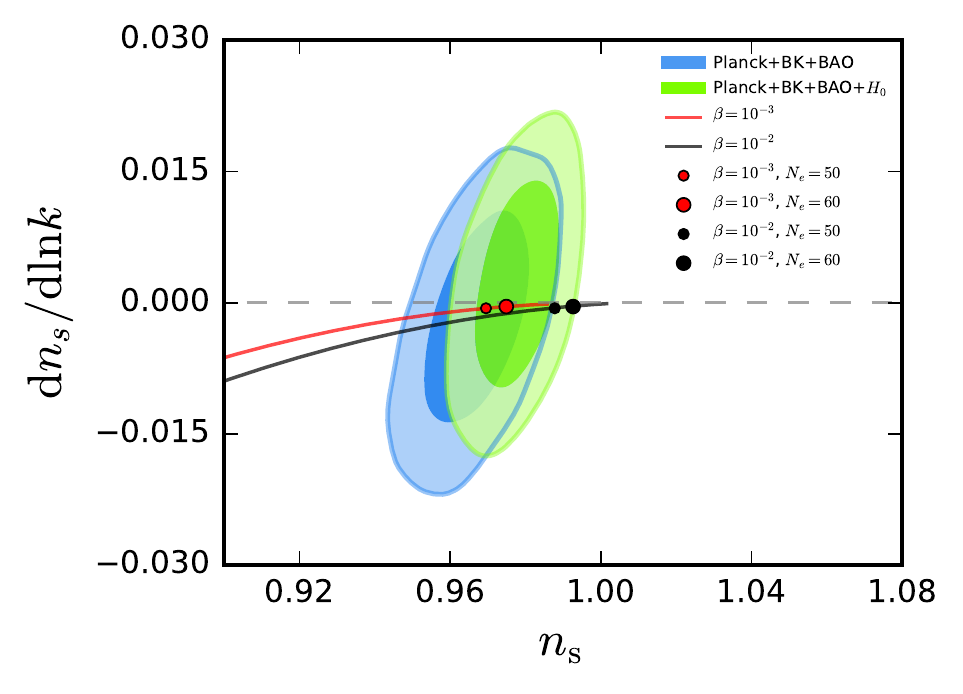}
\end{center}
\caption{Constraints ($1\sigma$ and $2\sigma$) on the $\Lambda$CDM+$r$+$d n_{\rm s}/d \ln k$+$N_{\rm eff}$ model in the $n_{\rm s}$--$r$ (\emph{left}) and $n_{\rm s}$--$d n_{\rm s}/d \ln k$ (\emph{right}) plane by using the Planck+BK+BAO and Planck+BK+BAO+$H_{0}$ data, compared to the theoretical predictions of the KKLMMT model. }
\label{fig2}
\end{figure*}

We plot the theoretical predictions for the parameters $n_{\rm s}$ and $r$ of the KKLMMT model in the left panel of Fig.~\ref{fig2}, where the observational constraints are still from the Planck+BK+BAO and Planck+BK+BAO+$H_{0}$ data. The red line is for the $N_{e}=50$ case and the black line is for the $N_{e}=60$ case. Different shapes of circle, triangle, and inverted triangle correspond to the different predictions of $\beta=10^{-3}$, $\beta=10^{-2}$, and $\beta=10^{-1}$, respectively. We figure out when $\beta> 0.018$, the inflaton potential turns to be convex, which is ruled out by the current observations. Actually, we see that the predictions of $\beta>10^{-2}$ are almost excluded by the observational data. Only the case of $\beta=10^{-2}$ is marginally favored by considering the $H_{0}$ measurement. Namely, when the local $H_{0}$ measurement is added to the Planck+BK+BAO data combination, the value of the parameter $\beta$ can be amplified to $10^{-2}$. Without the addition of the new $H_{0}$ measurement, the allowed value is lower. Compared with the results (with the ``WMAP9+" data) of Ref.~\cite{Ma:2013xma} , our results make the value of the parameter $\beta$ be amplified by an order of magnitude.

In the right panel of Fig.~\ref{fig2}, we plot the theoretical predictions for the parameters $n_{\rm s}$ and $d n_{\rm s}/d \ln k$ of the KKLMMT model. Given the above discussion, we focus on two cases of $\beta=10^{-3}$ (red line) and $\beta=10^{-2}$ (black line). We see that parts of the predictions for the two cases are always located inside the observationally allowed regions. We plot the cases of $N_{e}=50$ (small dots) and $N_{e}=60$ (big dots) on the two lines. We see that the predictions of $\beta=10^{-3}$ with a range of $N_{e}\in [50,60]$ are consistent with both two data sets, while the predictions of $\beta=10^{-2}$ with a range of $N_{e}\in [50,60]$ is only consistent with the Planck+BK+BAO+$H_{0}$ data. This result confirms that the range of the parameter $\beta$ can be amplified to ${\cal O}(10^{-2})$ provided that the direct $H_{0}$ measurement is considered.

\subsection* {3.3 The IR DBI model}

\begin{figure}[ht!]
\begin{center}
\includegraphics[width=8.0cm]{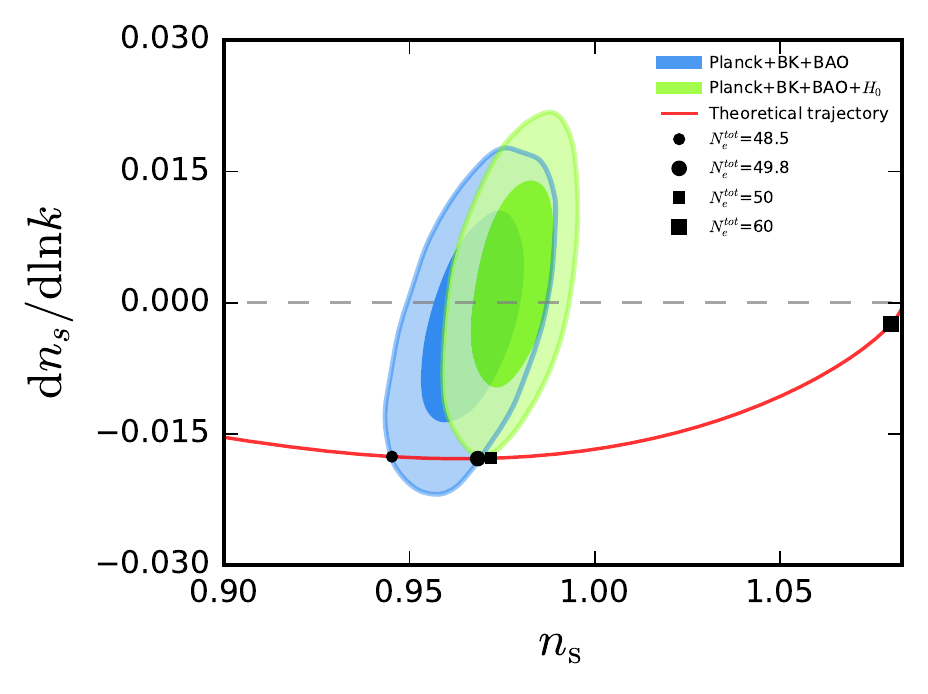}
\end{center}
\caption{Constraints ($1\sigma$ and $2\sigma$) on the $\Lambda$CDM+$r$+$d n_{\rm s}/d \ln k$+$N_{\rm eff}$ model in the $n_{\rm s}$--$d n_{\rm s}/d \ln k$ plane by using the Planck+BK+BAO and Planck+BK+BAO+$H_{0}$ data,  compared to the theoretical predictions of the IR DBI model. }
\label{fig3}
\end{figure}


The IR DBI model model can produce a somewhat large negative running. We plot the predicted trajectory (red line) in the parameter plane of $n_{\rm s}$--$d n_{\rm s}/d \ln k$ for the IR DBI model in Fig.~\ref{fig3}. We see that the predicted trajectory crosses the $2\sigma$ contour of the Planck+BK+BAO data, and is only tangent to the $2\sigma$ contour of the Planck+BK+BAO+$H_{0}$ data. For the case of $N_{e}^{\rm tot} \in [50,60]$, as shown by the trajectory between the two square dots in Fig.~\ref{fig3}, we find that the IR DBI model is excluded by the current observations at the 2$\sigma$ level. When smaller values of $N_{e}^{\rm tot}$ are considered, we see that the IR DBI model can be favored by the observations in some extent. We find that the range of $N_{e}^{\rm tot}\in [48.5,49.8]$ is consistent with the Planck+BK+BAO data, and the only case of $N_{e}^{\rm tot}=49.8$ is consistent with the Planck+BK+BAO+$H_{0}$ data. In addition, the IR DBI inflation also can only produce a small tensor mode, i.e., $r\ll 1/(N_{\rm e}^{\rm DBI^{4}}\gamma^{2})<10^{-6}$, which is undetectably small~\cite{Bean:2007eh}. We therefore do not show the $n_{\rm s}$--$r$ plot here because it provides no extra information.



To constrain the parameter $\beta$ of the IR DBI model, we further discuss the primordial non-Gaussianities. The Planck collaboration~\cite{Ade:2015ava} gives the constraints on the local, equilateral and orthogonal types of non-Gaussianity,
$$
\begin{aligned}
  &f_{NL}^{\rm loc}=0.8\pm5.0~(1\sigma),\\
  &f_{NL}^{\rm eq}=-4\pm43~(1\sigma),\\
  &f_{NL}^{\rm orth}=-26\pm21~(1\sigma),
\end{aligned}
$$
from the combined CMB temperature and polarization data. We compare the predictions of non-Gaussianities of the IR DBI model with the CMB observational results. The IR DBI model predicts an almost negligible local $f_{NL}$, which is totally consistent with the observational result. The orthogonal $f_{NL}$ satisfies $f_{NL}>0$ [see Eq.~(\ref{2.19})] in the IR DBI model, which is totally excluded by the observational results $f_{NL}^{\rm orth}<0$. Next we will have a look at the prediction for the equilateral type of non-Gaussianity in the IR DBI model.

\begin{figure}[ht!]
\begin{center}
\includegraphics[width=7.6cm]{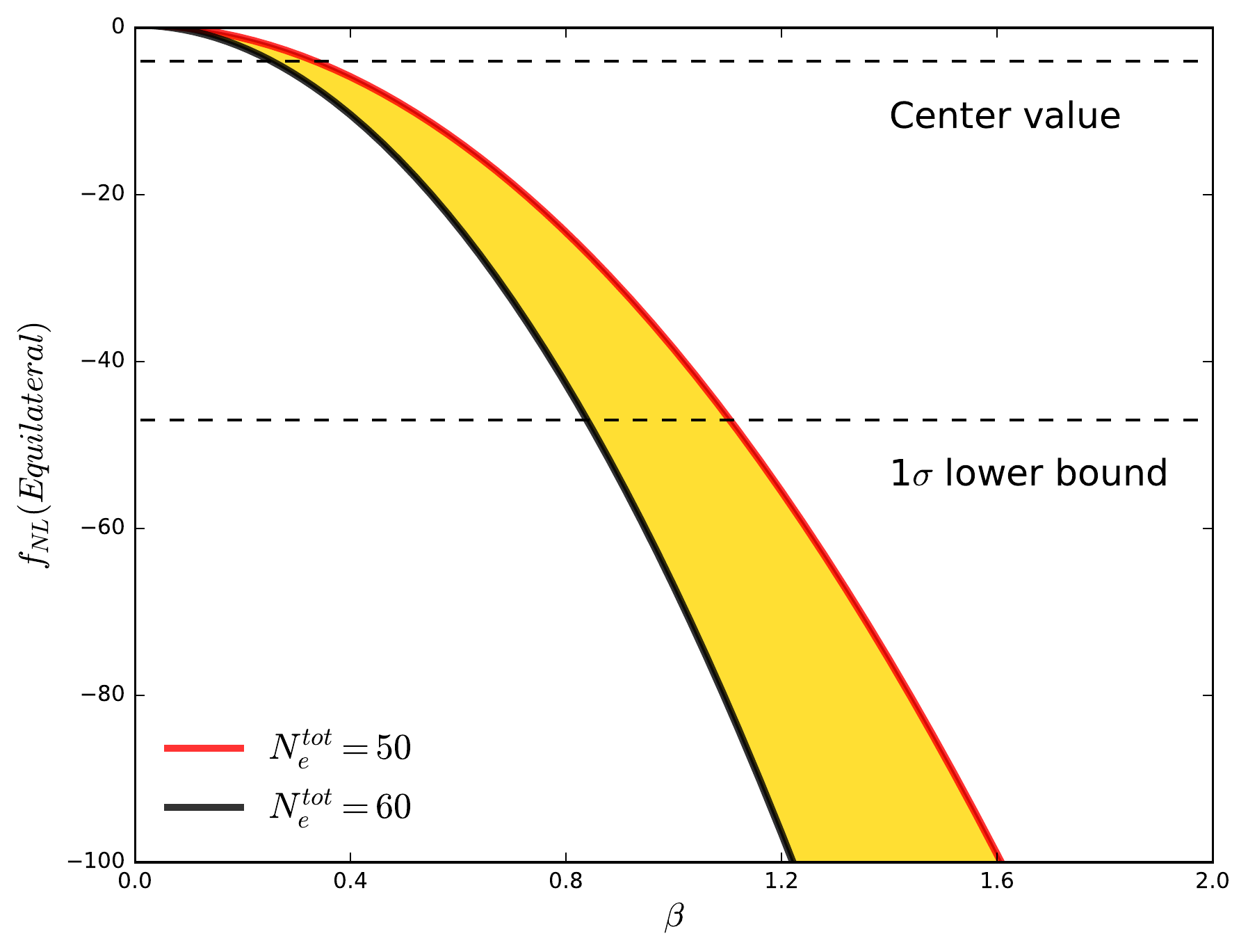}
\end{center}
\caption{Equilateral $f_{NL}$ predictions for different $\beta$ values in the IR DBI model. Here  we only plot the best-fit value and the lower bound ($1\sigma$) for $f_{NL}^{\rm eq}$ as dashed straight lines.}
\label{fig4}
\end{figure}

We give the equilateral non-Gaussianity $f_{NL}^{\rm eq}$, compared to the observational results in Fig.~\ref{fig4}. Still, we take the range of $N_{e}^{\rm tot}\in [50,60]$. In Fig.~\ref{fig4}, the yellow band corresponds to the predictions of the equilateral type of non-Gaussianity. The red line is for $N_{e}^{\rm tot}=50$ and the black line is for $N_{e}^{\rm tot}=60$. We find that, when $\beta <1.1$ ($1\sigma$), the predictions of the equilateral type of non-Gaussianity of the IR DBI model can fit the Planck TT, TE, EE+lowP data well. However, it should be pointed out that our settings for parameters of the IR DBI model are different from those in Ref.~\cite{Ade:2015lrj}. In Ref.~\cite{Ade:2015lrj}, by assuming a uniform prior $0.087 \leq c_{\rm s} \leq 1$, the results of $\beta<0.77$, $66<N_{e}^{\rm DBI}<72$, and $x<0.41$ ($2\sigma$) are obtained from the Planck TT+lowP data.

\section {Conclusion}\label{sec.4}

In this paper, we investigate the observational constraints on several typical brane inflation models by additionally considering the latest direct measurement of the Hubble constant. The brane inflation models discussed in this paper include a toy model (the prototype) of brane inflation, the KKLMMT model and the IR DBI model. We give the theoretical predictions of the above brane inflation models, and compare them with the constraint results of the $\Lambda$CDM+$r$+$d n_{\rm s}/d \ln k$+$N_{eff}$ model. We use the Planck+BK+BAO data combination and the Planck+BK+BAO+$H_{0}$ data combination in our analysis.

We find that, when the direct $H_{0}$ measurement is considered, the toy model of brane inflation is still in good agreement with the observational data at the $2\sigma$ level. For the KKLMMT model, the consideration of the new $H_{0}$ data allows the range of the parameter $\beta$ to be amplified to ${\cal O}(10^{-2})$ at the $2\sigma$ level; without adding the $H_{0}$ measurement, the allowed value of the parameter $\beta$ is required to be lower. This result indicates that the inclusion of the $H_{0}$ measurement makes the fine-tuning problem for the parameter $\beta$ be slightly relieved, which is what we expect for the KKLMMT model.

The IR DBI model predicts a larger negative running of spectral index than the toy model and the KKLMMT model. We find that the model can fit to the Planck+BK+BAO data when $N_{e}^{\rm tot} \in [48.5,49.8]$, but is only marginally favored by the Planck+BK+BAO+$H_{0}$ data. Namely, the latest $H_{0}$ measurement does not provide a better fit for the IR DBI model. Furthermore, we study the non-Gaussianities of the IR DBI model. The orthogonal non-Gaussianity in the IR DBI inflation is totally excluded by the current CMB data. We find that the parameter $\beta$ needs to be less than $1.1$ to guarantee the consistency for the equilateral non-Gaussianity between the theoretical prediction in the IR DBI model and the observational constraint from the current CMB data.


\section*{Acknowledgements}

This work was supported by the National Natural Science Foundation of China (Grants No.~11690021 and No.~11522540), the Top-Notch Young Talents Program of China, and the Provincial Department of Education of Liaoning (Grant No.~ L2012087).

\end{document}